\documentclass[sigconf,screen]{acmart}

\AtBeginDocument{%
  }

\setcopyright{acmlicensed}
\copyrightyear{2024}
\acmYear{2024}
\acmDOI{XXXXXXX.XXXXXXX}

\acmBooktitle{Companion Proceedings of the 33rd ACM Symposium on the Foundations of Software Engineering (FSE '25), June 23--27, 2025, Trondheim, Norway}
\acmISBN{978-1-4503-XXXX-X/18/06}

\usepackage{graphicx}
\usepackage{listings}
\usepackage{wrapfig}
\usepackage{caption}
\usepackage{booktabs}
\usepackage{tabularx}
\usepackage{float}
\usepackage{subcaption}
\usepackage{enumitem}
\usepackage{lscape}

\usepackage{multirow}
\usepackage{multicol}
\usepackage{url}
\usepackage{amsmath}
\usepackage{ragged2e}
\usepackage{color}
\usepackage{tabularray}
\usepackage{xcolor}
\usepackage{subfiles}

\begin{document}


 
 \title{From Documents to Dialogue: Building KG-RAG Enhanced AI Assistants}

\author{Manisha Mukherjee\textsuperscript{1}, Sungchul Kim\textsuperscript{2}, Xiang Chen\textsuperscript{2}, Dan Luo\textsuperscript{2}, Tong Yu\textsuperscript{2}, Tung Mai\textsuperscript{2}}

\affiliation{%
  \textsuperscript{1}Carnegie Mellon University, Pittsburgh, Pennsylvania, USA \\
  \textsuperscript{2}Adobe Research, San Jose, California \country{USA}
}


\renewcommand{\shortauthors}{Mukherjee et al.}

\begin{abstract}
The Adobe Experience Platform AI Assistant is a conversational tool that enables organizations to interact seamlessly with proprietary enterprise data through a chatbot. However, due to access restrictions, Large Language Models (LLMs) cannot retrieve these internal documents, limiting their ability to generate accurate zero-shot responses. To overcome this limitation, we use a Retrieval-Augmented Generation (RAG) framework powered by a Knowledge Graph (KG) to retrieve relevant information from external knowledge sources, enabling LLMs to answer questions over private or previously unseen document collections. In this paper, we propose a novel approach for building a high-quality, low-noise KG. We apply several techniques, including incremental entity resolution using seed concepts, similarity-based filtering to deduplicate entries, assigning confidence scores to entity-relation pairs to filter for high-confidence pairs, and linking facts to source documents for provenance. Our KG-RAG system retrieves relevant tuples, which are added to the user prompts context before being sent to the LLM generating the response. Our evaluation demonstrates that this approach significantly enhances response relevance, reducing \textit{irrelevant} answers by over \textbf{50\% } and increasing \textit{fully relevant} answers by \textbf{88\% } compared to the existing production system.

\end{abstract}


\keywords{Generative AI, LLM, KG, RAG, Knowledge Management}

\received{January 2025}

\maketitle

\section{Introduction}

\begin{figure}[h]
    
    \includegraphics[width=0.45\textwidth]{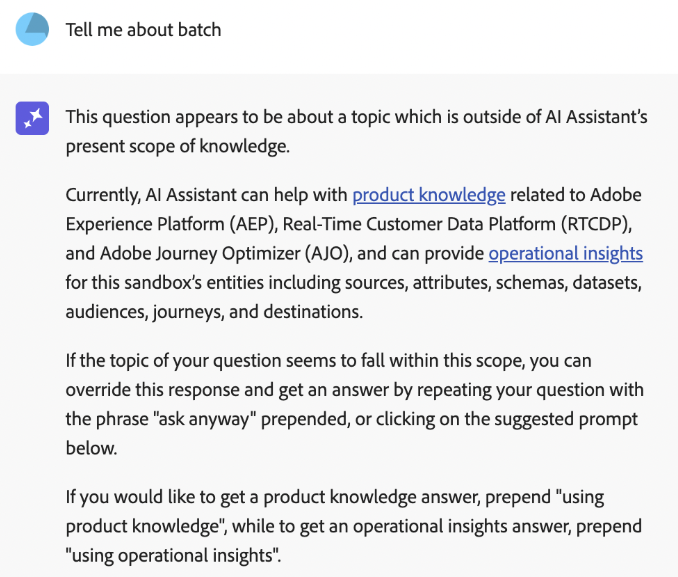} 
    \caption{Example of AI assistant generating a response saying that it is out of scope}
     \vspace{-1.5em}
    \label{fig:chat}
    
\end{figure}

Modern enterprises face a critical challenge: making their vast repositories of proprietary documentation and knowledge accessible to employees and customers. As organizations grow, they accumulate thousands of documents, including product specifications, user guides, internal processes, and customer support materials, all of which evolve over time. This institutional knowledge is valuable but often remains underutilized due to difficulties in quick and accurate information retrieval.

Consider a typical enterprise scenario: A customer support agent needs to quickly find specific information about a product feature, or a new employee needs to understand complex internal processes. Traditional search methods often fall short, returning either too many results or missing crucial contextual connections. This challenge is particularly acute in technology companies where product documentation is extensive and frequently updated.

The Adobe Experience Platform (AEP) is a customer data management and analytics solution that aggregates and analyzes data across various touchpoints. Integrated within AEP, the AI Assistant is a generative AI tool designed to assist users by providing insights into product functionality, operational data, and key business objects. Users interact with the AI Assistant by clicking on an icon in the upper right corner of the AEP user interface, which opens a right rail screen with a text box for entering prompts. While the AI Assistant aims to enhance productivity and support efficient navigation, the platform’s documentation spans over 17,000 pages, making it increasingly complex for users to locate precise and relevant information. Additionally, existing search systems lack semantic understanding, reducing their effectiveness for users seeking nuanced, contextually linked knowledge.


\begin{figure*}[h]
    
    \includegraphics[width=0.9\textwidth]{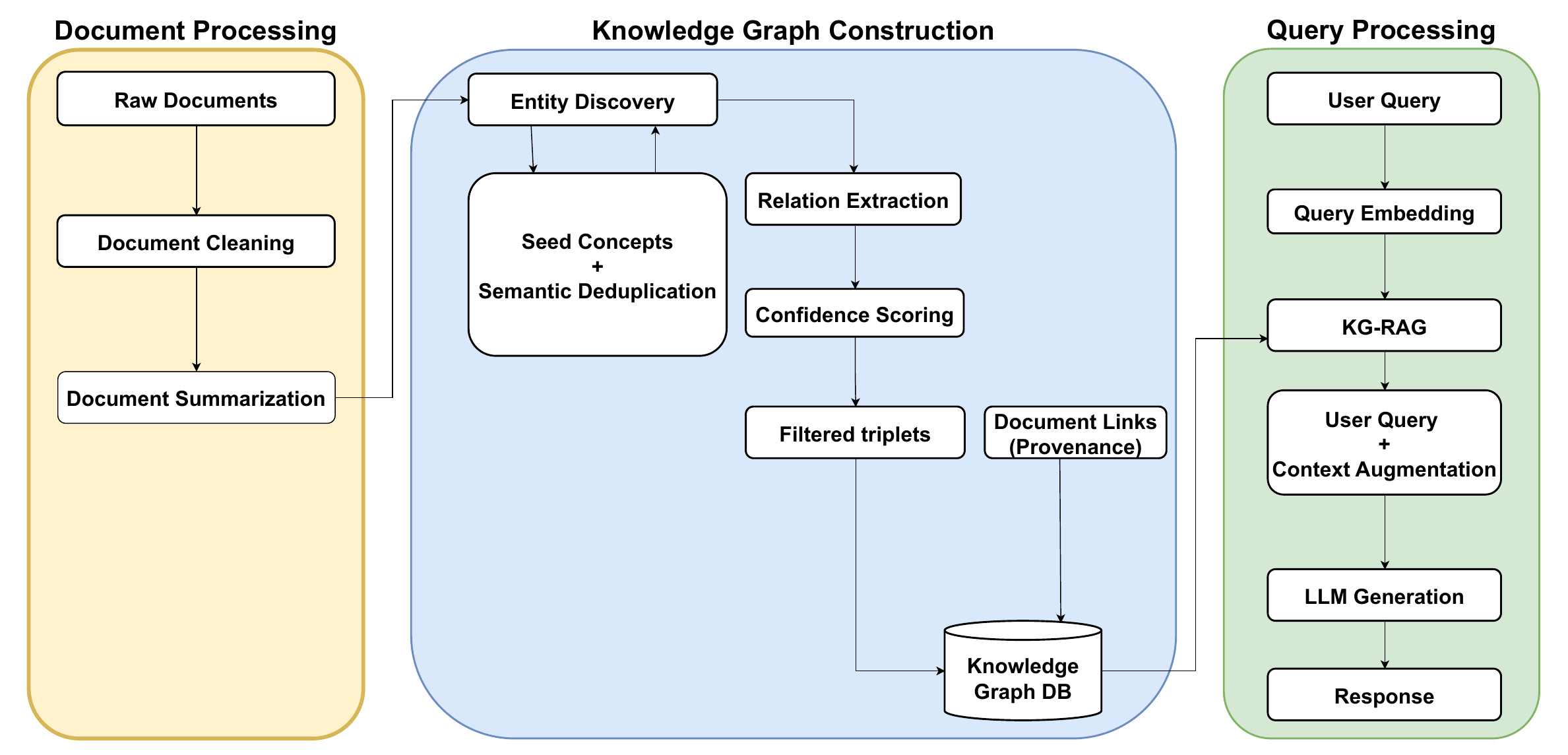} 
     \vspace{-1em}
    \caption{End-to-end pipeline for Knowledge Graph-based Retrieval-Augmented Generation (KG-RAG).}
     \vspace{-1em}
    \label{fig:kg_pipeline}
\end{figure*}

Large Language Models (LLMs) have become a promising solution for natural language interaction with enterprise knowledge bases, excelling at understanding user queries and generating human-like responses\cite{caldarini2022literature}. However, they face two key limitations in enterprise settings: first, LLMs trained on public data cannot access proprietary, organization-specific knowledge outside their training data; second,  without verified enterprise information, they may generate plausible but incorrect responses due to hallucination, potentially leading to confusion or errors in decision-making for unfamiliar users relying on these responses.

To address these limitations, Retrieval-Augmented Generation (RAG) has emerged as a paradigm that enhances LLMs by incorporating external knowledge sources at inference time \cite{gao2023retrieval}. However, current RAG implementations primarily rely on simple vector similarity search, which can miss important semantic relationships in complex enterprise documentation. Recent approaches also incorporate structured Knowledge Graphs (KGs) to improve semantic retrieval \cite{edge2024local}. However, KGs constructed using these techniques are often inconsistent and noisy due to unresolved and semantically duplicated entities and relations\cite{carta2023iterative}. These issues can reduce efficiency and accuracy, particularly in large-scale enterprise applications. In our work, we construct the KG incrementally, along with other techniques, to reduce noise, enabling more precise and contextually aware information retrieval for our AI Assistant.

The key contributions of our work include:
\begin{itemize}[noitemsep,topsep=0pt]
 
    \item A scalable pipeline for constructing enterprise KG incrementally from large document collections
    \item Novel techniques for maintaining graph quality through confidence scoring and semantic deduplication
    \item An integrated RAG system that leverages graph structure for improved response accuracy
    \item Empirical evaluation showing over 50\% reduction in irrelevant responses compared to traditional approaches
\end{itemize}

In the following sections, we detail our implementation approach (Section 2), evaluation methodology (Section 3), and results and discussion (Section 4), followed by a discussion of related work (Section 5) and conclusions (Section 6).

\section{Implementation}

We now unpack the high-level data flow of our pipeline (Figure \ref{fig:kg_pipeline}) for the system, referred to as KG-RAG throughout this paper, and provide implementation details for each step.

\subsection{Document cleaning and summarization}
In this phase, we focus on preparing the raw documents to make them suitable for entity extraction. This involves cleaning and summarizing the proprietary AEP document content from the 17,000 URLs collected. The first step in the document preparation pipeline is to clean the raw content extracted from these URLs by removing noise. For example, the original documents may contain HTML tags, images, videos, or scripts, which are removed to focus solely on the textual content. URLs and hyperlinks are identified and removed to avoid introducing irrelevant external references into the summary. Additionally, non-relevant metadata, such as author information, and footnotes, are excluded to maintain the contents relevance.

\begin{figure}[h]
    
    \includegraphics[width=0.45\textwidth]{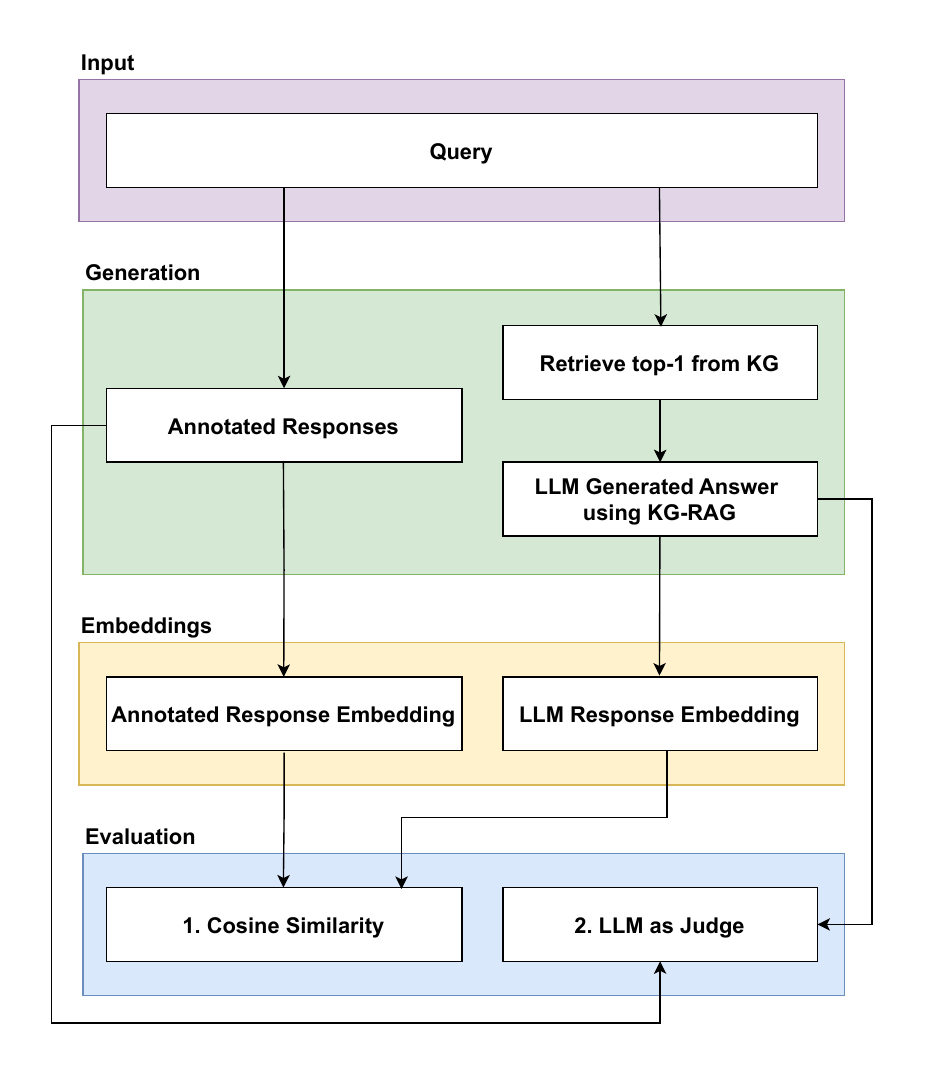} 
    \vspace{-2em}
    \caption{ Evaluation framework for response relevance. The pipeline compares baseline production answers with KG-RAG-enhanced responses.}
    \vspace{-2.0em}
    \label{fig:eval0}
\end{figure}

After cleaning the document, the next step is to generate a concise summary using an LLM. This reduces the document length, thus simplifying the document by highlighting key content and removing unnecessary details, aiding in the identification of relevant entities and relationships at the later stage.

\subsection{Seed Concept, Entity Discovery and Deduplication}

For each document, a summarized version was provided as input, along with a list of seed concepts or previously discovered entities (starting with an empty list). This list served as context for the LLM to identify new entities while avoiding redundancy. The LLM outputted a list of entities found in the document, including both new and known ones. 

A step was implemented to prevent semantic overlap, comparing newly identified entities to existing ones. A graph \( G \) can be described as \( G = (E, R) \), where \( E \) represents the collection of nodes (entities) and \( R \) denotes the set of edges (relations). Let \( e_i \) and \( e_j \) be two entities in \( E \). To improve the quality of the constructed KG, we impose the constraint that no two entities are semantically identical. Specifically, the following condition must hold:
 \vspace*{-1mm}
\begin{equation}
    \forall e_i, e_j \in E, \; i \neq j \quad 
    \label{overlap}
\end{equation}

If an entity was already in the KG (based on exact or close matches), it was excluded from being added again. As each new document was processed, the list was updated 
with any new, non-duplicate entities, ensuring the KG remained consistent and free from redundancy.

\subsection{Relation Extraction and Confidence Scoring}

After identifying the entities, the next step involves extracting the relationships between them, forming triplets. Once the list of entities has been identified in the previous step, it is sent, along with the current document to the LLM that then extracts potential relationships between the entities based on their contextual occurrence within the document. For instance, if the document contains the entities "segment" and "customer data," the model may extract a relationship such as "segment classifies customer data." Each triplet thus contributes to the construction of the KG, establishing connections between the entities.

Once the relationships are extracted, a confidence score is assigned by the LLM to each triplet to assess its reliability. This score ranges from 0 (low confidence) to 1 (high confidence), reflecting the certainty of the extracted relationship. For example, a triplet like (Batch processing, enhances, segmentation accuracy) may receive a high confidence score if it aligns with established industry practices, while a more ambiguous triplet such as (Batch processing, enhances, memory) might be assigned a lower confidence score due to its lack of clear association in the context. The confidence score allows us to filter the KG based on its value. In this paper, we use a threshold of 0.6 for confidence score for constructing the KG for the KG-RAG system. The threshold value was determined through extensive A/B testing.

\subsection{Provenance}

For each relationship in the KG, we maintain a record of its origin to ensure provenance. This is achieved by including the URL of the document from which the relationship was extracted. By incorporating provenance in this manner, we enable traceability, allowing the relationships within the KG to be verified and validated through reference to the original documents. While we do not specifically calculate how provenance influences query ranking, this feature primarily serves to allow users to interact with the source metadata and verify the origins of the information.

\section{Evaluation}

This section describes the evaluation framework used to measure the relevance of AI-generated answers. The process is summarized in the flowchart shown in Figure \ref{fig:eval0}.

\subsection{Dataset}

We use a dataset comprising 100 manually annotated question-answer pairs focused on the `segment' domain. These questions were collected based on production traffic, and the annotations were provided by domain experts at Adobe. The responses in the dataset were generated by the existing production system, also referred to as the baseline system, which retrieves the top documents based on the cosine similarity of embeddings and provides those documents, along with a prompt, to the LLM to generate the final answer. For each question, the answer generated by the AI assistant was presented to the annotators, who evaluated its relevance on a three-point scale: \textit{Fully Relevant} answers completely address the question,\textit{ Moderately Relevant} answers partially address the question but may lack completeness, and \textit{Irrelevant} answers fail to address the question meaningfully. To improve annotation consistency, detailed guidelines were provided to the annotators for scoring the responses. This dataset serves as a ground truth for evaluating the relevance and quality of generated answers using KG-RAG.

\subsection{LLM as Judge}

Manual annotation of responses is costly and time-consuming, making it challenging to scale evaluations effectively. To address this limitation, we utilize a LLM as a judge, offering a scalable and efficient alternative. Given the open-ended nature of answers, rule-based evaluation methods are impractical. Similarly, traditional text-matching metrics such as BLEU\cite{papineni2002bleu} and ROUGE\cite{lin2004rouge} are inadequate for assessing semantic relevance, as they rely heavily on surface-level word overlap rather than contextual meaning. The judge LLM evaluates relevance by interpreting the semantic content of responses, providing a more robust assessment aligned with human judgment.

\textbf{\textit{Prompt Design Strategies:}} To further enhance the judge LLM's performance on the evaluation task, we employ two strategies: 

\begin{itemize}
    \item In-context Learning (ICL) \cite{dong2022survey}: This method integrates task demonstrations into the prompt as illustrations. In our work, we randomly select 2-4 demonstrations from the training set, excluding these examples from the evaluation set.

    \item Chain-of-thought (CoT) \cite{wei2022chain}: This approach structures the input prompt to mimic human reasoning. To mitigate potential bias, chain-of-thought prompts were tested with diverse examples covering different semantic complexities. In our work, the judge model is required to generate an explanation first before providing the final judgment.
\end{itemize}

\begin{figure}[h]
    
    \includegraphics[width=0.45\textwidth]{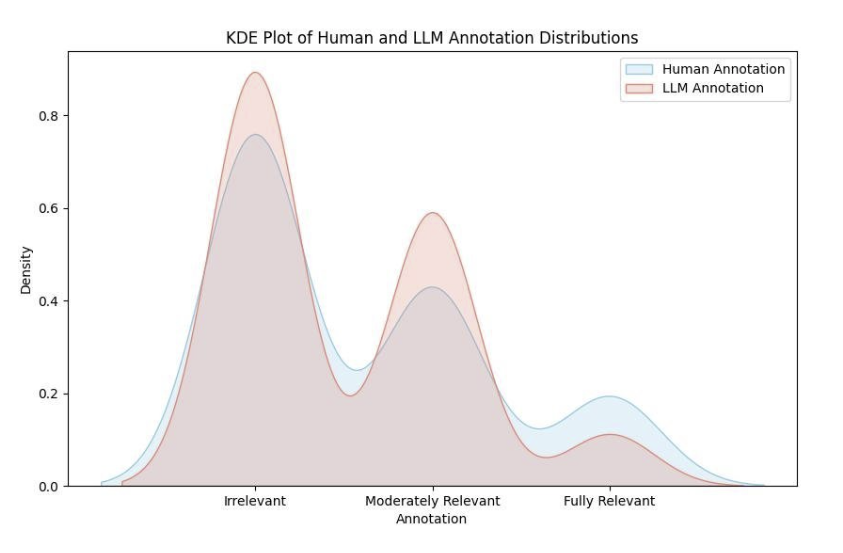} 
    \vspace{-2em}
    \caption{LLM-as-Judge KDE plot showing the distribution of relevance scores for baseline production and KG-RAG.}
    \vspace{-2.0em}
    \label{fig:kde}
\end{figure}

Our LLM as a judge aligns closely with human-annotated preferences, achieving \textbf{85\% }agreement. The Kernel Density Estimate (KDE) plot in Figure \ref{fig:kde} shows the distribution of relevance scores assigned by human evaluators and the LLM judge. The x-axis represents the relevance score on a scale from 0 to 3, where 1 indicates irrelevant answers and 3 indicates fully relevant answers. The y-axis shows the density or frequency of scores in that range. While the curves broadly overlap, indicating general agreement between human and LLM judgments, there is a some divergence at the upper end of the scale where the LLM curve shows less overlap with the human curve. When faced with ambiguous cases, the LLM appears to err on the side of caution by classifying more responses as `moderately relevant' compared to human evaluators. This conservative bias in the LLM judge actually helps prevent overestimation of system performance.

\subsection{Cosine Similarity for Semantic Matching}

In addition to LLM-based evaluation, we leverage cosine similarity \cite{gomaa2013survey} between sentence embeddings to measure semantic closeness between AI-generated answers and annotated ground-truth responses. Cosine similarity captures the contextual relationship between two vectors by measuring the cosine of the angle between them, making it invariant to the magnitude of the vectors and suitable for comparing textual content.

We generate embeddings for both the AI assistants answer without KG-RAG and answers after using KG-RAG using the bert-base-uncased model from HuggingFace \footnote{https://huggingface.co/google-bert/bert-base-uncased}. The cosine similarity between two vectors \( q \) and \( c \), representing the AI-assistant answers before and after incorporating KG-RAG, is defined as:
 \vspace*{-1mm}
\begin{equation}    
\text{Similarity}(q, c) = \cos(\theta) = \frac{\sum_{i=1}^{n} q_i c_i}{\sqrt{\sum_{i=1}^{n} q_i^2} \sqrt{\sum_{i=1}^{n} c_i^2}}
\label{cosine}
\end{equation}

Here, \( q_i \) represents the embedding of the AI-assistant answer of the baseline production, and \( c_i \) represents the embedding of the answer with KG-RAG. The cosine similarity provides a value between -1 and 1, where 1 indicates identical semantic content, 0 indicates no similarity, and -1 indicates complete dissimilarity.


This combination of LLM-based scoring and cosine similarity offers a comprehensive and scalable evaluation framework that balances qualitative judgment with quantitative metrics.

\section{Results and Discussion}
Our proposed KG-RAG approach was evaluated against the existing production system as a baseline. The production system retrieves the top documents based on the cosine similarity of embeddings, then provides those documents, along with the user prompt, to the LLM to generate the final answer. Adobe's agnostic approach to LLMs allows us to select the best-in-class technology for the task at hand. Currently, the AI Assistant in AEP\cite{adobe_ai_assistant_2025} leverages Microsoft's Azure OpenAI Service to answer queries.

The results demonstrate a significant improvement in answer relevance with the integration of the KG-RAG retriever. For the results shown in Table \ref{tab:results}, we retrieved the top-1 (top-k, where k=1) tuple from the KG for each query and appended it as additional context to the user query before passing it to the generative model. The model-generated response was then evaluated by an LLM judge, categorizing answers into three relevance types: \textit{Irrelevant}, \textit{Moderately Relevant}, and \textit{Fully Relevant}. Table \ref{tab:results} compares the performance of the baseline with the KG-RAG system based on these relevance annotations.

The baseline production system, produced a substantial number of irrelevant answers (52 instances). By integrating the KG-RAG system, irrelevant responses were reduced by 51.9\%, highlighting the effectiveness of KG retrieval in filtering noise and aligning content with user queries. Moderately relevant responses increased by 71\%. Most notably, fully relevant responses nearly doubled, from 17 to 32, marking an 88.2\% improvement.

In addition to evaluating using LLM as a judge, we also computed the average cosine similarity between answers generated by KG-RAG and baseline snippets. The KG-RAG system achieved an average cosine similarity of 0.89 with the \textit{fully relevant} snippets. This high score indicates greater semantic overlap, reinforcing the KG-RAG approach's ability to generate answers that more closely resemble annotated correct answers in meaning, beyond surface-level wording.

Despite its advantages, the implementation faced several challenges. Balancing noise reduction and entity coverage was challenging. Scalability and latency were also concerns, particularly when processing large document collections. Ensuring the quality of the KG required regular validation, automated checks, and periodic human review to prevent the generation of a noisy or inconsistent KG.
\begin{table}[htbp]
\centering
\resizebox{\columnwidth}{!}{
\renewcommand{\arraystretch}{1.3}
\begin{tabular}{l|c|c|c}
\hline
\textbf{Response Category} & \textbf{Baseline} & \textbf{KG-RAG} & \textbf{Improvement (\%)} \\
\hline\hline
Irrelevant & 52 & 25 & $-$51.9 \\
Moderately Relevant & 31 & 53 & $+$71.0 \\
Fully Relevant & 17 & 32 & $+$88.2 \\
\hline
\end{tabular}
}

\caption{Comparison of response relevance between baseline and KG-RAG systems. The baseline system retrieves the top documents based on the cosine similarity of embeddings and provides those documents, along with a prompt, to the LLM to generate the final answer.}
\vspace{-3.75em}
\label{tab:results}
\end{table}

Although the system was developed within Adobe's ecosystem, the approach should be adaptable to other enterprise contexts. The document processing pipeline, KG construction, and entity resolution techniques can be customized to fit different domains. Furthermore, the KG-RAG system can be integrated with various LLM providers and adjusted to meet specific performance and retrieval requirements.

\textit{\textbf{Limitations:}}  
While the data used in this study is real-world and collected from production traffic, it may not fully capture the diversity of business data, and sandbox testing may not fully reflect the complexities of live production systems. The small dataset, though relevant, limits generalizability, and future work will focus on expanding it to enhance robustness. Additionally, the use of LLMs as evaluators introduces potential biases and variability. To mitigate this, multiple evaluation methods were employed to cross-validate results, and extensive tuning was performed to align LLM judgments with human-annotated preferences, achieving 85\% agreement. Due to proprietary restrictions, the prompts and code cannot be released; however, detailed methodology descriptions are provided to ensure reproducibility.

\textit{\textbf{Future Work:}} We aim to improve the accuracy of entity and relation extraction by utilizing fine-tuned models for embedding generation. We also plan to explore top-k retrieval to complement the reliance on the top-1 result, which could enhance coverage and increase the likelihood of retrieving the most relevant answers. These improvements will aim to refine accuracy and broaden the applicability of our approach, supporting more robust and scalable KG based solutions for answering user queries.

\section{Related Work}

LLMs have shown remarkable performance improvements across tasks such as machine translation \cite{xu2024contrastive}, text summarization \cite{jin2024comprehensive}, and question answering \cite{saito2024unsupervised}. However, their limitations in accessing proprietary or dynamically updated data without hallucinating errors remain a significant barrier. Several approaches attempt to mitigate these issues by integrating domain-specific knowledge into language models \cite{agrawal2023can,mukherjee2023stack}.

RAG frameworks have gained traction in combining language models with external knowledge sources\cite{lewis2020retrieval,salemi2024evaluating}. The basic principle involves retrieving relevant data from external documents and integrating it into the context of a language model at inference time. Early implementations of RAG relied on vector similarity search to embed and retrieve text chunks\cite{gao2023retrieval}. These methods improved semantic search efficiency but often struggled with complex reasoning due to the lack of structured data representation.

Recent advancements in leveraging structured Knowledge Graphs (KGs) for RAG have provided promising alternatives. \citeauthor{edge2024local}\cite{edge2024local} introduced a graph-enhanced retrieval mechanism, demonstrating that structured data linked through relationships can significantly improve contextual relevance in query responses. Unlike simple vector-based retrieval, KGs offer explicit semantic connections between concepts, allowing richer contextual augmentation for natural language understanding.

Many studies now utilize LLMs for constructing KGs \cite{li2024contextualization}, focusing on tasks such as named entity recognition and classification \cite{li2021span,zhou2023universalner}, relation extraction \cite{jiang2024genres}. Complementary work by \citeauthor{shu2024knowledge}\cite{shu2024knowledge} highlights the role of LLMs in link prediction and reasoning over KGs, suggesting that hybrid approaches outperform purely generative models.

Our implementation of KG-RAG incorporates several concepts from related systems. For example, \cite{lairgi2024itext2kg} builds the knowledge graph without duplication but lacks features such as confidence scores and provenance. In contrast, \cite{amaral2022prove} includes provenance but does not incorporate the other features, while \cite{edge2024local} assigns confidence scores but also performs community detection on the graph and generates summaries for each community using LLMs. Our work aligns with this trajectory by advancing the integration of knowledge graphs into RAG systems tailored for enterprise documentation. Specifically, we enhance the KG with incremental entity resolution using seed concepts, similarity-based filtering to deduplicate entries, assigning confidence scores to entity-relation pairs to prioritize high-confidence pairs, and linking facts to source documents for provenance.

In terms of evaluation methodologies, \cite{zheng2023judging} emphasize the importance of combining LLM-based judges with other metrics such as cosine similarity for assessing response relevance. While cosine similarity provides a quantitative measure of semantic overlap, LLM evaluators interpret nuanced meanings that static metrics miss. Our evaluation framework adopts a hybrid approach, complementing automated scoring with human-like judgments to balance scalability and interpretive depth.

\section{Conclusion}
In this paper, we present an enterprise implementation for incremental KG construction. Our method incrementally builds the KG, improving the precision of entity and relation extraction while addressing challenges such as deduplication and provenance. We also demonstrate how our KG-RAG based retriever can be integrated with an LLM to provide more accurate and contextually relevant answers to user queries. Initial evaluations demonstrate that our method significantly improves performance over the existing production baseline by reducing \textit{irrelevant} answers by more than 50\% and increasing \textit{fully relevant} answers by 88\%, based on a manually annotated dataset. Additionally, an average cosine similarity of 0.89 was observed between \textit{fully relevan}t answers produced by the baseline system and those generated by the KG-RAG-based system. These findings highlight the effectiveness of our approach and provide a promising framework for leveraging KGs containing verified, up-to-date facts from documents that may not have been part of the LLM's pretraining data. This KG-RAG retrieval system represents one component of the broader Adobe AI Assistant, which integrates additional features to further enhance response relevance and the overall user experience.

\bibliographystyle{ACM-Reference-Format}
\bibliography{sample-base}


\begin{thebibliography}{25}


\ifx \showCODEN    \undefined \def \showCODEN     #1{\unskip}     \fi
\ifx \showDOI      \undefined \def \showDOI       #1{#1}\fi
\ifx \showISBNx    \undefined \def \showISBNx     #1{\unskip}     \fi
\ifx \showISBNxiii \undefined \def \showISBNxiii  #1{\unskip}     \fi
\ifx \showISSN     \undefined \def \showISSN      #1{\unskip}     \fi
\ifx \showLCCN     \undefined \def \showLCCN      #1{\unskip}     \fi
\ifx \shownote     \undefined \def \shownote      #1{#1}          \fi
\ifx \showarticletitle \undefined \def \showarticletitle #1{#1}   \fi
\ifx \showURL      \undefined \def \showURL       {\relax}        \fi
\providecommand\bibfield[2]{#2}
\providecommand\bibinfo[2]{#2}
\providecommand\natexlab[1]{#1}
\providecommand\showeprint[2][]{arXiv:#2}

\bibitem[{Adobe}(2025)]%
        {adobe_ai_assistant_2025}
\bibfield{author}{\bibinfo{person}{{Adobe}}.} \bibinfo{year}{2025}\natexlab{}.
\newblock \bibinfo{title}{Adobe AI Assistant in AEP Security Fact Sheet}.
\newblock
\newblock
\urldef\tempurl%
\url{https://www.adobe.com/content/dam/cc/en/trust-center/ungated/whitepapers/experience-cloud/adobe-ai-assistant-in-aep-security-fact-sheet.pdf}
\showURL{%
\tempurl}
\newblock
\shownote{Accessed: 2025-01-16}.


\bibitem[Agrawal et~al\mbox{.}(2023)]%
        {agrawal2023can}
\bibfield{author}{\bibinfo{person}{Garima Agrawal}, \bibinfo{person}{Tharindu Kumarage}, \bibinfo{person}{Zeyad Alghamdi}, {and} \bibinfo{person}{Huan Liu}.} \bibinfo{year}{2023}\natexlab{}.
\newblock \showarticletitle{Can knowledge graphs reduce hallucinations in llms?: A survey}.
\newblock \bibinfo{journal}{\emph{arXiv preprint arXiv:2311.07914}} (\bibinfo{year}{2023}).
\newblock


\bibitem[Amaral et~al\mbox{.}(2022)]%
        {amaral2022prove}
\bibfield{author}{\bibinfo{person}{Gabriel Amaral}, \bibinfo{person}{Odinaldo Rodrigues}, {and} \bibinfo{person}{Elena Simperl}.} \bibinfo{year}{2022}\natexlab{}.
\newblock \showarticletitle{ProVe: A pipeline for automated provenance verification of knowledge graphs against textual sources}.
\newblock \bibinfo{journal}{\emph{Semantic Web}} \bibinfo{number}{Preprint} (\bibinfo{year}{2022}), \bibinfo{pages}{1--34}.
\newblock


\bibitem[Caldarini et~al\mbox{.}(2022)]%
        {caldarini2022literature}
\bibfield{author}{\bibinfo{person}{Guendalina Caldarini}, \bibinfo{person}{Sardar Jaf}, {and} \bibinfo{person}{Kenneth McGarry}.} \bibinfo{year}{2022}\natexlab{}.
\newblock \showarticletitle{A literature survey of recent advances in chatbots}.
\newblock \bibinfo{journal}{\emph{Information}} \bibinfo{volume}{13}, \bibinfo{number}{1} (\bibinfo{year}{2022}), \bibinfo{pages}{41}.
\newblock


\bibitem[Carta et~al\mbox{.}(2023)]%
        {carta2023iterative}
\bibfield{author}{\bibinfo{person}{Salvatore Carta}, \bibinfo{person}{Alessandro Giuliani}, \bibinfo{person}{Leonardo Piano}, \bibinfo{person}{Alessandro~Sebastian Podda}, \bibinfo{person}{Livio Pompianu}, {and} \bibinfo{person}{Sandro~Gabriele Tiddia}.} \bibinfo{year}{2023}\natexlab{}.
\newblock \showarticletitle{Iterative zero-shot llm prompting for knowledge graph construction}.
\newblock \bibinfo{journal}{\emph{arXiv preprint arXiv:2307.01128}} (\bibinfo{year}{2023}).
\newblock


\bibitem[Dong et~al\mbox{.}(2022)]%
        {dong2022survey}
\bibfield{author}{\bibinfo{person}{Qingxiu Dong}, \bibinfo{person}{Lei Li}, \bibinfo{person}{Damai Dai}, \bibinfo{person}{Ce Zheng}, \bibinfo{person}{Jingyuan Ma}, \bibinfo{person}{Rui Li}, \bibinfo{person}{Heming Xia}, \bibinfo{person}{Jingjing Xu}, \bibinfo{person}{Zhiyong Wu}, \bibinfo{person}{Tianyu Liu}, {et~al\mbox{.}}} \bibinfo{year}{2022}\natexlab{}.
\newblock \showarticletitle{A survey on in-context learning}.
\newblock \bibinfo{journal}{\emph{arXiv preprint arXiv:2301.00234}} (\bibinfo{year}{2022}).
\newblock


\bibitem[Edge et~al\mbox{.}(2024)]%
        {edge2024local}
\bibfield{author}{\bibinfo{person}{Darren Edge}, \bibinfo{person}{Ha Trinh}, \bibinfo{person}{Newman Cheng}, \bibinfo{person}{Joshua Bradley}, \bibinfo{person}{Alex Chao}, \bibinfo{person}{Apurva Mody}, \bibinfo{person}{Steven Truitt}, {and} \bibinfo{person}{Jonathan Larson}.} \bibinfo{year}{2024}\natexlab{}.
\newblock \showarticletitle{From local to global: A graph rag approach to query-focused summarization}.
\newblock \bibinfo{journal}{\emph{arXiv preprint arXiv:2404.16130}} (\bibinfo{year}{2024}).
\newblock


\bibitem[Gao et~al\mbox{.}(2023)]%
        {gao2023retrieval}
\bibfield{author}{\bibinfo{person}{Yunfan Gao}, \bibinfo{person}{Yun Xiong}, \bibinfo{person}{Xinyu Gao}, \bibinfo{person}{Kangxiang Jia}, \bibinfo{person}{Jinliu Pan}, \bibinfo{person}{Yuxi Bi}, \bibinfo{person}{Yi Dai}, \bibinfo{person}{Jiawei Sun}, {and} \bibinfo{person}{Haofen Wang}.} \bibinfo{year}{2023}\natexlab{}.
\newblock \showarticletitle{Retrieval-augmented generation for large language models: A survey}.
\newblock \bibinfo{journal}{\emph{arXiv preprint arXiv:2312.10997}} (\bibinfo{year}{2023}).
\newblock


\bibitem[Gomaa et~al\mbox{.}(2013)]%
        {gomaa2013survey}
\bibfield{author}{\bibinfo{person}{Wael~H Gomaa}, \bibinfo{person}{Aly~A Fahmy}, {et~al\mbox{.}}} \bibinfo{year}{2013}\natexlab{}.
\newblock \showarticletitle{A survey of text similarity approaches}.
\newblock \bibinfo{journal}{\emph{international journal of Computer Applications}} \bibinfo{volume}{68}, \bibinfo{number}{13} (\bibinfo{year}{2013}), \bibinfo{pages}{13--18}.
\newblock


\bibitem[Jiang et~al\mbox{.}(2024)]%
        {jiang2024genres}
\bibfield{author}{\bibinfo{person}{Pengcheng Jiang}, \bibinfo{person}{Jiacheng Lin}, \bibinfo{person}{Zifeng Wang}, \bibinfo{person}{Jimeng Sun}, {and} \bibinfo{person}{Jiawei Han}.} \bibinfo{year}{2024}\natexlab{}.
\newblock \showarticletitle{GenRES: Rethinking Evaluation for Generative Relation Extraction in the Era of Large Language Models}.
\newblock \bibinfo{journal}{\emph{arXiv preprint arXiv:2402.10744}} (\bibinfo{year}{2024}).
\newblock


\bibitem[Jin et~al\mbox{.}(2024)]%
        {jin2024comprehensive}
\bibfield{author}{\bibinfo{person}{Hanlei Jin}, \bibinfo{person}{Yang Zhang}, \bibinfo{person}{Dan Meng}, \bibinfo{person}{Jun Wang}, {and} \bibinfo{person}{Jinghua Tan}.} \bibinfo{year}{2024}\natexlab{}.
\newblock \showarticletitle{A comprehensive survey on process-oriented automatic text summarization with exploration of llm-based methods}.
\newblock \bibinfo{journal}{\emph{arXiv preprint arXiv:2403.02901}} (\bibinfo{year}{2024}).
\newblock


\bibitem[Lairgi et~al\mbox{.}(2024)]%
        {lairgi2024itext2kg}
\bibfield{author}{\bibinfo{person}{Yassir Lairgi}, \bibinfo{person}{Ludovic Moncla}, \bibinfo{person}{R{\'e}my Cazabet}, \bibinfo{person}{Khalid Benabdeslem}, {and} \bibinfo{person}{Pierre Cl{\'e}au}.} \bibinfo{year}{2024}\natexlab{}.
\newblock \showarticletitle{itext2kg: Incremental knowledge graphs construction using large language models}. In \bibinfo{booktitle}{\emph{International Conference on Web Information Systems Engineering}}. Springer, \bibinfo{pages}{214--229}.
\newblock


\bibitem[Lewis et~al\mbox{.}(2020)]%
        {lewis2020retrieval}
\bibfield{author}{\bibinfo{person}{Patrick Lewis}, \bibinfo{person}{Ethan Perez}, \bibinfo{person}{Aleksandra Piktus}, \bibinfo{person}{Fabio Petroni}, \bibinfo{person}{Vladimir Karpukhin}, \bibinfo{person}{Naman Goyal}, \bibinfo{person}{Heinrich K{\"u}ttler}, \bibinfo{person}{Mike Lewis}, \bibinfo{person}{Wen-tau Yih}, \bibinfo{person}{Tim Rockt{\"a}schel}, {et~al\mbox{.}}} \bibinfo{year}{2020}\natexlab{}.
\newblock \showarticletitle{Retrieval-augmented generation for knowledge-intensive nlp tasks}.
\newblock \bibinfo{journal}{\emph{Advances in Neural Information Processing Systems}}  \bibinfo{volume}{33} (\bibinfo{year}{2020}), \bibinfo{pages}{9459--9474}.
\newblock


\bibitem[Li et~al\mbox{.}(2024)]%
        {li2024contextualization}
\bibfield{author}{\bibinfo{person}{Dawei Li}, \bibinfo{person}{Zhen Tan}, \bibinfo{person}{Tianlong Chen}, {and} \bibinfo{person}{Huan Liu}.} \bibinfo{year}{2024}\natexlab{}.
\newblock \showarticletitle{Contextualization distillation from large language model for knowledge graph completion}.
\newblock \bibinfo{journal}{\emph{arXiv preprint arXiv:2402.01729}} (\bibinfo{year}{2024}).
\newblock


\bibitem[Li et~al\mbox{.}(2021)]%
        {li2021span}
\bibfield{author}{\bibinfo{person}{Fei Li}, \bibinfo{person}{ZhiChao Lin}, \bibinfo{person}{Meishan Zhang}, {and} \bibinfo{person}{Donghong Ji}.} \bibinfo{year}{2021}\natexlab{}.
\newblock \showarticletitle{A span-based model for joint overlapped and discontinuous named entity recognition}.
\newblock \bibinfo{journal}{\emph{arXiv preprint arXiv:2106.14373}} (\bibinfo{year}{2021}).
\newblock


\bibitem[Lin(2004)]%
        {lin2004rouge}
\bibfield{author}{\bibinfo{person}{Chin-Yew Lin}.} \bibinfo{year}{2004}\natexlab{}.
\newblock \showarticletitle{Rouge: A package for automatic evaluation of summaries}. In \bibinfo{booktitle}{\emph{Text summarization branches out}}. \bibinfo{pages}{74--81}.
\newblock


\bibitem[Mukherjee and Hellendoorn(2023)]%
        {mukherjee2023stack}
\bibfield{author}{\bibinfo{person}{Manisha Mukherjee} {and} \bibinfo{person}{Vincent~J Hellendoorn}.} \bibinfo{year}{2023}\natexlab{}.
\newblock \showarticletitle{Stack over-flowing with results: the case for domain-specific pre-training over one-size-fits-all models}.
\newblock \bibinfo{journal}{\emph{Parameters (billion)}} \bibinfo{volume}{10}, \bibinfo{number}{1} (\bibinfo{year}{2023}), \bibinfo{pages}{100}.
\newblock


\bibitem[Papineni et~al\mbox{.}(2002)]%
        {papineni2002bleu}
\bibfield{author}{\bibinfo{person}{Kishore Papineni}, \bibinfo{person}{Salim Roukos}, \bibinfo{person}{Todd Ward}, {and} \bibinfo{person}{Wei-Jing Zhu}.} \bibinfo{year}{2002}\natexlab{}.
\newblock \showarticletitle{Bleu: a method for automatic evaluation of machine translation}. In \bibinfo{booktitle}{\emph{Proceedings of the 40th annual meeting of the Association for Computational Linguistics}}. \bibinfo{pages}{311--318}.
\newblock


\bibitem[Saito et~al\mbox{.}(2024)]%
        {saito2024unsupervised}
\bibfield{author}{\bibinfo{person}{Kuniaki Saito}, \bibinfo{person}{Kihyuk Sohn}, \bibinfo{person}{Chen-Yu Lee}, {and} \bibinfo{person}{Yoshitaka Ushiku}.} \bibinfo{year}{2024}\natexlab{}.
\newblock \showarticletitle{Unsupervised llm adaptation for question answering}.
\newblock \bibinfo{journal}{\emph{arXiv preprint arXiv:2402.12170}} (\bibinfo{year}{2024}).
\newblock


\bibitem[Salemi and Zamani(2024)]%
        {salemi2024evaluating}
\bibfield{author}{\bibinfo{person}{Alireza Salemi} {and} \bibinfo{person}{Hamed Zamani}.} \bibinfo{year}{2024}\natexlab{}.
\newblock \showarticletitle{Evaluating retrieval quality in retrieval-augmented generation}. In \bibinfo{booktitle}{\emph{Proceedings of the 47th International ACM SIGIR Conference on Research and Development in Information Retrieval}}. \bibinfo{pages}{2395--2400}.
\newblock


\bibitem[Shu et~al\mbox{.}(2024)]%
        {shu2024knowledge}
\bibfield{author}{\bibinfo{person}{Dong Shu}, \bibinfo{person}{Tianle Chen}, \bibinfo{person}{Mingyu Jin}, \bibinfo{person}{Chong Zhang}, \bibinfo{person}{Mengnan Du}, {and} \bibinfo{person}{Yongfeng Zhang}.} \bibinfo{year}{2024}\natexlab{}.
\newblock \showarticletitle{Knowledge graph large language model (KG-LLM) for link prediction}.
\newblock \bibinfo{journal}{\emph{arXiv preprint arXiv:2403.07311}} (\bibinfo{year}{2024}).
\newblock


\bibitem[Wei et~al\mbox{.}(2022)]%
        {wei2022chain}
\bibfield{author}{\bibinfo{person}{Jason Wei}, \bibinfo{person}{Xuezhi Wang}, \bibinfo{person}{Dale Schuurmans}, \bibinfo{person}{Maarten Bosma}, \bibinfo{person}{Fei Xia}, \bibinfo{person}{Ed Chi}, \bibinfo{person}{Quoc~V Le}, \bibinfo{person}{Denny Zhou}, {et~al\mbox{.}}} \bibinfo{year}{2022}\natexlab{}.
\newblock \showarticletitle{Chain-of-thought prompting elicits reasoning in large language models}.
\newblock \bibinfo{journal}{\emph{Advances in neural information processing systems}}  \bibinfo{volume}{35} (\bibinfo{year}{2022}), \bibinfo{pages}{24824--24837}.
\newblock


\bibitem[Xu et~al\mbox{.}(2024)]%
        {xu2024contrastive}
\bibfield{author}{\bibinfo{person}{Haoran Xu}, \bibinfo{person}{Amr Sharaf}, \bibinfo{person}{Yunmo Chen}, \bibinfo{person}{Weiting Tan}, \bibinfo{person}{Lingfeng Shen}, \bibinfo{person}{Benjamin Van~Durme}, \bibinfo{person}{Kenton Murray}, {and} \bibinfo{person}{Young~Jin Kim}.} \bibinfo{year}{2024}\natexlab{}.
\newblock \showarticletitle{Contrastive preference optimization: Pushing the boundaries of llm performance in machine translation}.
\newblock \bibinfo{journal}{\emph{arXiv preprint arXiv:2401.08417}} (\bibinfo{year}{2024}).
\newblock


\bibitem[Zheng et~al\mbox{.}(2023)]%
        {zheng2023judging}
\bibfield{author}{\bibinfo{person}{Lianmin Zheng}, \bibinfo{person}{Wei-Lin Chiang}, \bibinfo{person}{Ying Sheng}, \bibinfo{person}{Siyuan Zhuang}, \bibinfo{person}{Zhanghao Wu}, \bibinfo{person}{Yonghao Zhuang}, \bibinfo{person}{Zi Lin}, \bibinfo{person}{Zhuohan Li}, \bibinfo{person}{Dacheng Li}, \bibinfo{person}{Eric Xing}, {et~al\mbox{.}}} \bibinfo{year}{2023}\natexlab{}.
\newblock \showarticletitle{Judging llm-as-a-judge with mt-bench and chatbot arena}.
\newblock \bibinfo{journal}{\emph{Advances in Neural Information Processing Systems}}  \bibinfo{volume}{36} (\bibinfo{year}{2023}), \bibinfo{pages}{46595--46623}.
\newblock


\bibitem[Zhou et~al\mbox{.}(2023)]%
        {zhou2023universalner}
\bibfield{author}{\bibinfo{person}{Wenxuan Zhou}, \bibinfo{person}{Sheng Zhang}, \bibinfo{person}{Yu Gu}, \bibinfo{person}{Muhao Chen}, {and} \bibinfo{person}{Hoifung Poon}.} \bibinfo{year}{2023}\natexlab{}.
\newblock \showarticletitle{Universalner: Targeted distillation from large language models for open named entity recognition}.
\newblock \bibinfo{journal}{\emph{arXiv preprint arXiv:2308.03279}} (\bibinfo{year}{2023}).
\newblock


\end{thebibliography}
\end{document}